\documentclass[aps,twocolumn,showpacs]{revtex4}
\pdfoutput=1
\usepackage{graphicx}
\usepackage{amsmath}
\usepackage{amssymb}
\usepackage{hyperref}
\usepackage{latexsym}
\usepackage{epsfig}


\begin{document}

\title{Long-lived dipolar molecules and Feshbach molecules in a 3D optical lattice}

\author{Amodsen Chotia,$^{\ast}$ Brian Neyenhuis,$^{\ast}$ Steven A. Moses, Bo Yan, Jacob P. Covey, Michael Foss-Feig, Ana Maria Rey, Deborah S. Jin,$^\dagger$ and Jun Ye$^\dagger$}
\affiliation{
\normalsize{JILA, National Institute of Standards and Technology and University of Colorado, Department of Physics, University of Colorado, Boulder, CO 80309-0440, USA}
}

\begin{abstract}
We have realized long-lived ground-state polar molecules in a 3D optical lattice, with a lifetime of up to 25~s, which is limited only by off-resonant scattering of the trapping light. Starting from a 2D optical lattice, we observe that the lifetime increases dramatically as a small lattice potential is added along the tube-shaped lattice traps. The 3D optical lattice also
dramatically increases the lifetime for weakly bound Feshbach molecules. For a pure gas of Feshbach molecules, we observe a lifetime of $>$20 s in a 3D optical lattice; this represents a 100-fold improvement over previous results. This lifetime is also limited by off-resonant scattering, the rate of which is related to the size of the Feshbach molecule. Individually trapped Feshbach molecules in the 3D lattice can be converted to pairs of K and Rb atoms and back with nearly 100$\%$ efficiency.

\end{abstract}


\maketitle

Controllable long-range and anisotropic dipole-dipole interactions can enable novel applications of quantum gases in investigating strongly correlated many-body systems~\cite{Baranov200871, Pupillo, 2009RPPh...72l6401L, PhysRevLett.107.115301, 2009NJPh...11e5049C, RevModPhys.80.885}.
Recent experiments have realized an ultracold gas of polar molecules in the ro-vibrational ground state~\cite{2008Sci...322..231N} with high-resolution, single-state control at the level of hyperfine structure~\cite{2010PhRvL.104c0402O}.  However, an obstacle to creating long-lived quantum gases of polar molecules was encountered with the observation of bimolecular chemical reactions in the quantum regime~\cite{2010Sci...327..853O}.  Even with the demonstrated strong suppression of the reaction rate for spin-polarized fermionic KRb molecules, the lifetime of a 300 nK sample with a peak density of 10$^{12}$/cm$^3$ was limited to $\sim$1 s.  Furthermore, when an external electric field is applied to polarize the molecules in the lab frame, the attractive part of the dipole-dipole interaction dramatically increases the ultracold chemical reaction rate, reducing the lifetime of the dipolar gas to a few ms when the lab-frame molecular dipole moment reaches 0.2 Debye~\cite{2010Natur.464.1324N}.  A promising recent development was the demonstration that confining fermionic polar molecules in a 1D optical lattice suppresses the rate of chemical reactions even in the presence of dipolar interactions.  Here, the spatial anisotropy of the dipolar interaction was exploited by confining a gas of oriented KRb molecules in a two-dimensional geometry to suppress the attractive part of the dipolar interaction and thus achieve control of the stereodynamics of the bimolecular reactions~\cite{2011NatPh...7..502D}. In this regime, the lifetime of a trapped gas of polar molecules with a lab-frame dipole moment  of $\sim$0.2 Debye, a temperature of 800 nK, and a number density of 10$^{7}$ cm$^{-2}$, was $\sim$1~s.

In this letter, we study KRb molecules confined in 2D and 3D optical lattice traps, where we explore the effects of the lattice confinement on the lifetime of the ultracold gas.  We note that a lifetime of 8~s has been achieved for homonuclear Cs$_2$ molecules in a 3D lattice~\cite{2010NatPh...6..265D}.  In our work, we find that long lifetimes are achieved for the molecules in a strong 3D lattice trap, even when there is a significant dipole moment in the lab frame.  In addition, we observe that adding a weak axial corrugation to a 2D lattice can result in long lifetimes for the trapped molecules.

The experiments start with an ultracold mixture of $2.9 \times 10^5$ $^{40}$K atoms and $2.3 \times 10^5$ $^{87}$Rb atoms in a crossed optical dipole trap (ODT) at 1064 nm, at a temperature that is twice the Rb condensation temperature $T_c$.  The trap frequencies for Rb are $21$ Hz in the horizontal ($x,y$) plane and $165$ Hz in the vertical ($z$) direction; the trap frequencies for K are 1.37 times larger. The atoms are transferred into a 3D optical lattice in three steps. We first turn on a retro-reflected vertical beam in $150$ ms to create a weak 1D lattice. In the second step, the ODT is ramped off in $100$ ms so that the two beams used for the ODT (which propagate along $x$ and $y$) can be converted to lattice beams by allowing them to be retro-reflected. The intensities along all three directions are then ramped to their final values in $100$ ms. The three lattice beams are derived from a common laser but individually frequency shifted.
The $x$ and $y$ beams are elliptical with a 200 $\times$ 40 $\mu$m waist and are linearly polarized orthogonal in the $x$-$y$ plane; the $z$-beam has a circular waist of 250 $\mu$m and is linearly polarized along $x$.  We calibrate the lattice strength using Rb atoms with two different methods (Kapitza-Dirac scattering pulse in a BEC \cite{KapitzaDirac} and parametric modulation of the lattice) and then account for the differences in mass and ac polarizability to determine the lattice strength for KRb molecules.
The values reported herein for the lattice depth in each direction are expressed in units of the molecule recoil energy, $E_R$, and have an estimated $10\%$ uncertainty.

Once the atoms are loaded in the 3D lattice, we ramp an external magnetic field across an $s$-wave Feshbach resonance at $546.78$ G to form loosely bound $^{40}$K$^{87}$Rb molecules with an efficiency of about 10$\%$.  With the Feshbach molecules at $B=545.8$ G, where their binding energy is $h\times$400 kHz, we use two-photon stimulated Raman adiabatic passage (STIRAP) to coherently transfer the Feshbach molecules to the ro-vibrational ground state~\cite{2008Sci...322..231N}, with a typical one-way transfer efficiency of 80\%. All the molecules are in a single nuclear spin state in the rotational ground state, $|N=0, m_N=0, m_I^K=-4, m_I^{Rb}=1/2\rangle$, following the notation defined in~\cite{2010PhRvL.104c0402O}.
During this procedure, unpaired K and Rb atoms are removed using resonant light pulses. To measure the number of ground-state molecules in the lattice, we reverse the STIRAP process and then image the resultant Feshbach molecules using absorption of a probe beam that is tuned to the imaging transition for K atoms.

\begin{figure}[htbp]
			\centering
						\includegraphics[width=0.47\textwidth]{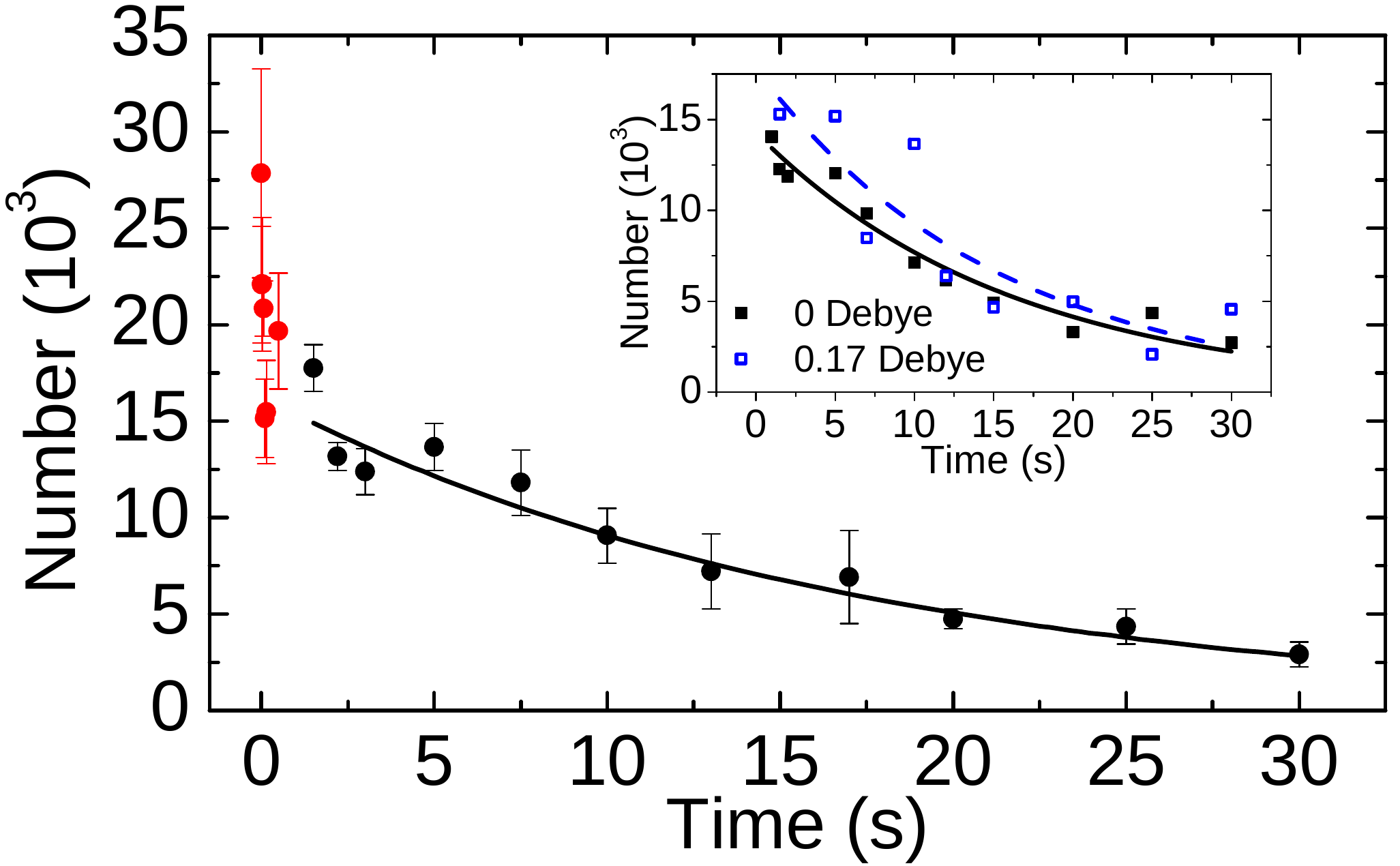}
				\caption{Loss of ground-state KRb molecules as a function of time in a 3D lattice with depths of 56, 56, and 70 $E_R$ in $x$, $y$, and $z$, respectively, where $E_R=\hbar^2 k^2/2m$ is the KRb recoil energy, $k$ is the magnitude of the lattice beam wave vector, and $m$ the molecular mass. Neglecting the very short time points (red solid circles), the number of molecules for times larger than 1~s (black solid circles) are fit to a single exponential decay, yielding a 1/$e$ lifetime of 16.3$\pm$1.5 s. Inset: Lifetime in an isotropic lattice with a depth of 50 $E_R$, with (blue open squares, $0.17$ Debye) and without (black squares, 0 Debye) an applied electric field. The lifetimes at $0.17$ Debye (15$\pm$4 s) and 0 Debye agree within uncertainty.}
			\label{GSM2}
		\end{figure}

\begin{figure}[htbp]
			\centering
						\includegraphics[width=0.47\textwidth]{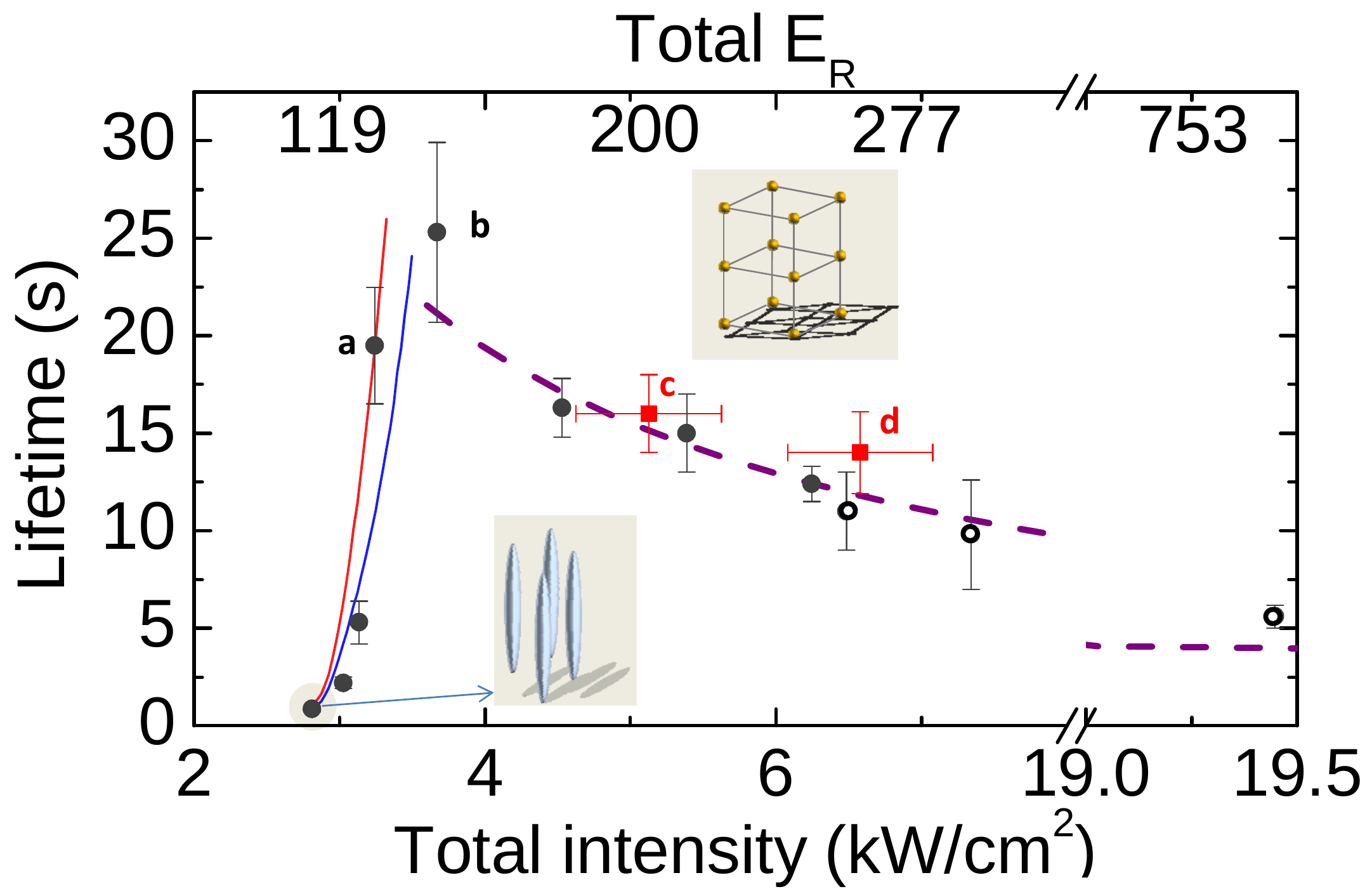}
				\caption{Lifetime of KRb ground-state molecules in an optical lattice. Black circles: $x$ and $y$ lattice beams are fixed at 56~$E_R$ per beam, while $z$ is varied from 0 to 136 $E_R$ (1 $E_R$ corresponds to a lattice intensity $I$ = $0.025$ kW/cm$^2$). The lifetime reaches a maximum of 25$\pm$5~s when the $z$ lattice depth is 34 $E_R$ (point ~\textbf{b}). For higher lattice intensities, the lifetime decreases, which is consistent with loss due to off-resonant light scattering (dashed line). The open circles correspond to a 3D lattice where the radial confinement was also varied. The red squares correspond to lifetimes measured with an additional traveling-wave beam at 1064~nm illuminating the molecules in the 3D lattice. Point~\textbf{c} (\textbf{d}) corresponds to the 3D lattice of point~\textbf{a} with an intensity of 3.2 kW/cm$^2$ (\textbf{b} with 3.7 kW/cm$^2$) plus the additional beam intensity of 2.3 kW/cm$^2$ (3.5 kW/cm$^2$). Solid lines: see text.}
			\label{GSM}
		\end{figure}

Figure~\ref{GSM2} shows a time-dependent evolution of the ground-state molecule population in the 3D lattice. In the first few 100's of ms, the measured number of molecules exhibits relatively large variations in repeated iterations of the experiment and is consistent with some fast initial decay.  In all our measurements of ground-state molecules in deep 3D lattices (for example, in the data for Fig.~\ref{GSM}), we observe a similar feature.  One possible explanation for this fast decay is collisions of the ground-state molecules with impurities, such as molecules in excited internal states that might be produced in the STIRAP process. Fitting the data for times greater than 1~s to an exponential decay, which is consistent with a single-body loss mechanism, gives a $1/e$ lifetime of 16.3$\pm$1.5~s.  This is much longer than previously measured lifetimes of trapped ultracold polar molecules of about 1~s in an ODT~\cite{2010Sci...327..853O} or in a 1D lattice~\cite{2011NatPh...7..502D}.

The long lifetime for ground-state molecules in a reasonably deep 3D lattice can be understood simply from the fact that the optical lattice localizes the molecules and therefore prevents bimolecular reactions. It was previously seen that an applied electric field strongly increased the chemical reaction rate~\cite{2010Natur.464.1324N}.  However, for molecules individually isolated in a 3D lattice, we expect no dependence of the lifetime on the strength of an applied electric field.  In the inset to Fig.~\ref{GSM2}, we show that indeed we do not observe any decrease of the lifetime for polarized molecules with an induced dipole moment of $0.17$ Debye.

To understand what limits the lifetime of the molecules in the 3D lattice, we investigate its dependence on the lattice strength as summarized in Fig.~\ref{GSM}. First, we explore the transition from a 2D lattice (an array of one-dimensional tubes) to a 3D lattice.
For a molecular gas confined in the tubes with no lattice in $z$, we find a lifetime of $\sim$1~s. However, as soon as a small lattice potential is added along $z$, the lifetime is dramatically increased, reaching 5~s at 12 $E_R$ and 20~s at $17$$E_R$ (point~\textbf{a} in Fig.~\ref{GSM}). To verify that bimolecular reactions are the dominant loss mechanism, we have checked that the lifetime in uncorrugated tubes decreases significantly (to 0.1 s) when we apply an electric field (oriented along the tubes) that gives an induced dipole moment of 0.17 Debye. In addition, the fact that we can place an upper limit of 10\% of the initial number remaining at long times puts a limit on the contribution to our signal from tubes that are occupied with only one molecule.

We consider a number of factors (Supplementary Information) to understand the rapid suppression of loss as a function of the lattice strength in $z$. In general, Pauli blocking for identical fermions and dissipation blockade effects (suppression of loss when the loss rate for particles on the same site is much larger than the tunneling rate \cite{2008Sci...320.1329S}) can play a role in the lifetime of KRb molecules in an optical lattice.  However, in the measurements reported here, the optical lattice is sparsely filled.  For our identical fermionic molecules, we expect Pauli blocking would give an even steeper function than we observe. Moreover, for a 5~$E_R$ lattice in $z$, the lifetime of KRb molecules in the tube does not change significantly in the presence of an applied electric field. This observation suggests that an incoherent process, such as heating of the trapped gas, limits the lifetime in this regime of weak lattice confinement. Instabilities in the optical phase of the lattice beams directly give rise to translational noise of the lattice, which can promote molecules to higher bands, where they have increased mobility and could then collide with other molecules. A simple theoretical model taking into account a constant heating rate, with collisions in higher bands happening on a timescale much shorter than the heating time, is consistent with the experimental observation (red and blue solid lines in Fig.~\ref{GSM}, with heating rates of 1~$E_R$/s (66 nK/s) and 2~$E_R$/s, respectively).

In Fig.~\ref{GSM}, the lifetime reaches a maximum of 25$\pm$5~s indicated by point~\textbf{b}. As the intensity is increased further, the lifetime starts to decrease, consistent with off-resonant photon scattering becoming the dominant loss mechanism. The rich internal state structure of molecules ensures that each off-resonant photon scattering event has a high probability of causing the loss of a molecule from the ground state. To explore this effect, we added an additional traveling-wave beam with a wavelength of 1064 nm; this increases the photon scattering rate without increasing the trap depth and we observe a significant reduction of the lifetime due to the additional light. We can extract the imaginary part of the polarizability of KRb molecules by fitting the lifetime as a function of the light intensity to $1/(\alpha I)$. Here, $\alpha$ is the imaginary part of the polarizability at 1064~nm, which we determine to be $(2.052\pm0.009) \times 10^{-12}$ MHz/(W/cm$^2$); this is consistent with a theory estimate for KRb~\cite{SKotochigova2011,2009NJPh...11e5043K}.

We have also explored the lifetime of KRb Feshbach molecules in the 3D optical lattice. It has been shown that these weakly bound molecules can be rapidly lost from an ODT due to collisions with atoms~\cite{PhysRevLett.100.143201}. Even with removal of the Rb atoms and the RF transfer of the K atoms to a different hyperfine state, all previously measured lifetimes for KRb Feshbach molecules were less than 10 ms~\cite{2008NatPh...4..622O}. However, with the ability to create ground-state molecules, which do not scatter light that is resonant with the single-atom transitions, we can more efficiently use light pulses to remove any residual atoms. A few ms after the atom removal, we reverse the STIRAP process and recreate a Feshbach molecule gas. For the case of molecules in the ODT (no lattice), we find that this extends the lifetime of the Feshbach molecules to 150 ms.  Therefore, we can conclude that previous lifetime measurements were likely limited by collisions with residual atoms.

\begin{figure}[htbp]
			\centering
						\includegraphics[width=0.47\textwidth]{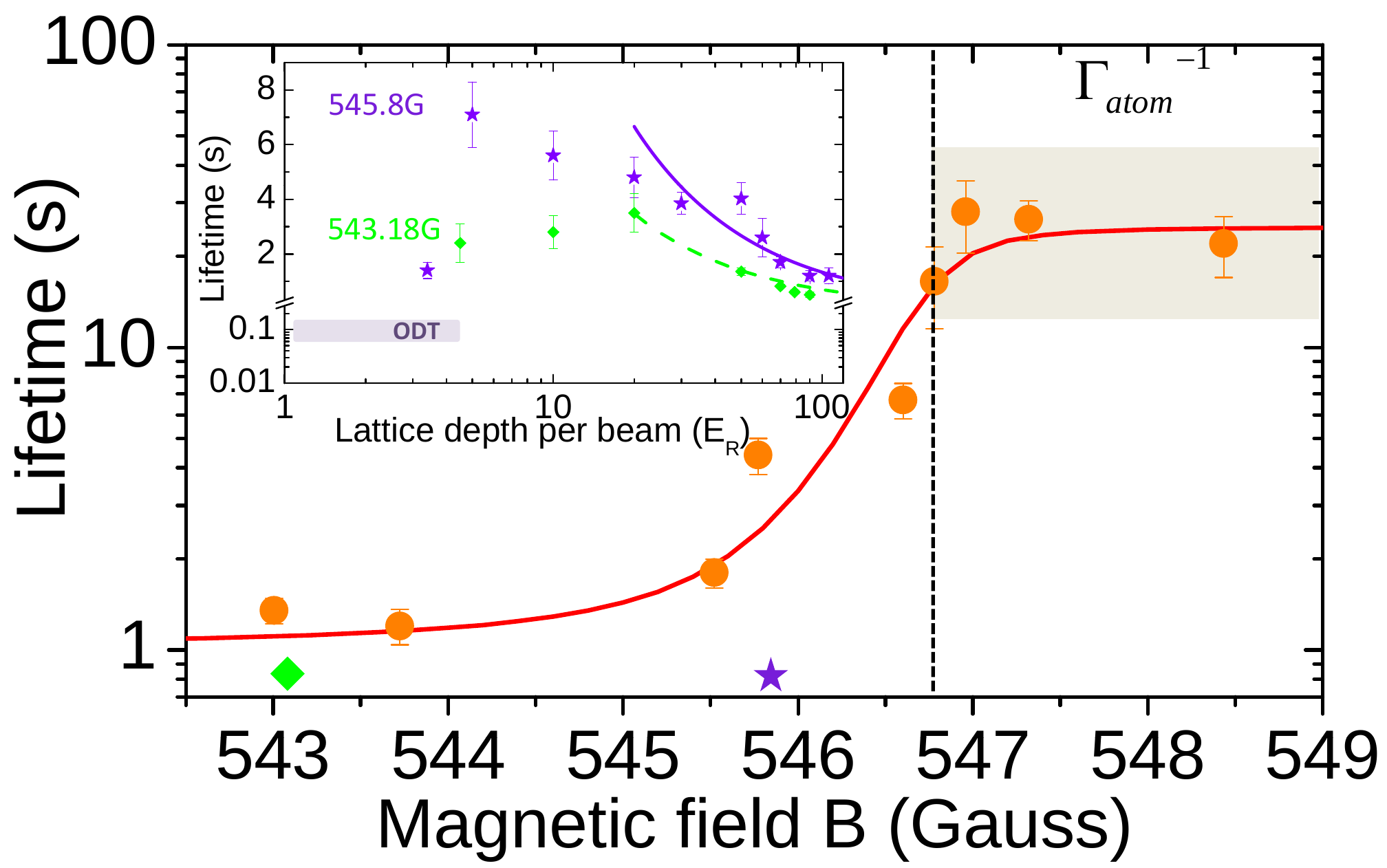}
				\caption{Lifetime of Feshbach molecules and confinement-induced molecules measured as a function of $B$.
A purified sample of Feshbach molecules is held in an isotropic 3D optical lattice (50~$E_R$ per beam, 20~kHz trap frequency).  Near the Feshbach resonance, the loss rate due to photon scattering can be modeled (solid line) as a weighted sum of the free atom loss rate $\Gamma_{atom}$ and a higher loss rate for tightly bound molecules $\Gamma_{molecule}$ . The grey shaded area indicates the single atom lifetime, and its uncertainty, measured for the same experimental conditions. Inset: Lifetime of Feshbach molecules in a 3D lattice as a function of the trap intensity, at 545.8~G (blue stars) and 543.18~G (green diamonds). The dashed and solid line fits are used to extract the scattering rates.}
			\label{FBM}
		\end{figure}

When we perform the procedure described above for KRb molecules in an optical lattice, we find that the purified gas of Feshbach molecules can have a lifetime as long as 10~s. In Fig.~\ref{FBM} we explore the lifetime of the Feshbach molecules in the 3D lattice as a function of the magnetic-field detuning from the resonance ($B_0$ = 546.78~G), where varying the magnetic field, $B$, changes the binding energy and the size of the Feshbach molecules.  We start by forming a purified sample of the Feshbach molecules in a strong 3D lattice with an intensity of 50~$E_R$ per beam at $B$ = 545.8~G. The magnetic field is then ramped to its final value in 1~ms. At the end of the hold time in the 3D lattice, $B$ is ramped back to 545.8~G where we image the molecules.

Above the Feshbach resonance, the lattice potential allows for the existence of confinement-induced molecules that do not exist in free space~\cite{PhysRevLett.96.030401}. We find that confinement-induced molecules have a lifetime (25~s) that is comparable to that for K or Rb atoms in the same trap. Below the Feshbach resonance, the molecule lifetime decreases quickly when the magnetic field is ramped to lower values. Several Gauss below the resonance, the Feshbach molecule lifetime is reduced to $\sim$1~s, which is still significantly longer than in the ODT.  Overall, these results represent a two-orders-of-magnitude improvement in the lifetime of Feshbach molecules compared to a previous measurement of KRb molecules in an optical lattice~\cite{2006PhRvL..97l0402O}.

To understand the dependence of the lifetime on $B$, we can assume that the lifetime is limited by off-resonant photon scattering from the lattice light and consider two limiting cases. For $B \gg B_0$, the photon scattering limit is simply that for free atoms $\Gamma _{atom}$; for $B \ll B_0$, we have a higher photon scattering rate $\Gamma _{molecule}$ due to a larger wavefunction overlap with electronically excited molecules~\cite{2009NJPh...11e5036D}.  In a two-channel model of the Feshbach resonance~\cite{RevModPhys.82.1225},  the Feshbach molecule wavefunction can be written as an amplitude $Z^{1/2}$ times the bare ``closed-channel''  molecule wavefunction plus an amplitude $(1 - Z)^{1/2}$ times the ``open channel'' wavefunction that describes the scattering state of two free atoms. We then take the total photon scattering rate to be given by $Z\Gamma _{molecule} + (1 - Z)\Gamma _{atom}$. With pairs of atoms confined in an optical trap with a known depth, $Z$ can be calculated straightforwardly with a coupled-channel theory~\cite{RevModPhys.82.1225,B820917K}. Using the measured loss rates for the limiting cases, $\Gamma_{atom}$ and $\Gamma_{molecule}$, this simple theory (solid line in Fig.~\ref{FBM}) without any additional adjustable parameters describes very well the experimental results (filled circles).  We note that the rate of atom-molecule collisions has been analyzed in a similar way~\cite{2008PhRvA..77c2726D,2006PhRvL..97l0402O}.

The assumption that the lifetime of the purified gas of Feshbach molecules in a 3D lattice is limited by only the photon scattering can be checked by varying the lattice beam intensity. This is shown in the inset of Fig.~\ref{FBM} for two values of $B$. Similar to the case of ground-state molecules, we observe a rapid initial increase in the lifetime going from no lattice (only the ODT) to a weak lattice. Following this initial rise, we observe a decrease in the Feshbach molecule lifetime as the lattice intensity is increased, consistent with loss due to off-resonant scattering of the lattice light. From the fits of Fig.~\ref{FBM} we extract an imaginary part of the Feshbach molecule's ac polarizability at 1064~nm of 15.9$\pm$1.6 MHz/(W/cm$^2$) for $B$ = 545.8~G and 30$\pm$3 MHz/(W/cm$^2$) for $B$ = 543.18~G.

In a simple model for the conversion of atoms to Feshbach molecules in a 3D lattice, one could assume 100$\%$ conversion efficiency for individual lattice sites that are occupied by exactly one K atom and one Rb atom.
Starting with a purified sample of Feshbach molecules prepared with round-trip STIRAP and atom removal as discussed above, we dissociate the molecules in the lattice by ramping $B$ above $B_0$ to 548.97~G. This should ideally produce only pre-formed pairs of atoms. We then ramp $B$ back down to 545.8~G and measure the molecular conversion efficiency.  The result is $(87\pm13)\%$, where the uncertainty is dominated by fluctuations in STIRAP efficiency in successive runs of the experiment. This high efficiency is far above the maximum of $25\%$ observed in an ODT~\cite{2008PhRvA..78a3416Z}; this indicates that optimizing the loading procedure in order to have a larger number of sites with exactly one Rb and one K atom would increase the overall conversion efficiency. For heteronuclear Bose-Fermi mixtures, optimizing the number of pre-formed atom pairs in a lattice remains a challenge. Recent progress in this direction includes the characterization of a dual Bose-Fermi Mott insulator with two isotopes of Yb~\cite{2011NatPhSugawa}, as well as proposals to use interaction effects to optimize the lattice loading~\cite{PhysRevLett.89.040402, PhysRevLett.90.110401, PhysRevA.81.011605}.

The capability demonstrated here for using a 3D lattice to freeze out chemical reactions, and thus prepare an ensemble of long-lived dipolar molecules, opens the door for studying many-body interactions in a gas of polar molecules in a lattice. For example, spectroscopy of rotational states of polar molecules in a lattice is a possible approach to access correlation functions~\cite{ PhysRevA.84.033608}. While individual molecules may experience long rotational coherence times, dipolar interactions between neighboring lattice sites, which could have an interaction energy on the order of a few hundred Hz, will certainly require a systematic understanding of the many-body system to understand the resulting complex response.  Future challenges in studying these systems include achieving higher lattice filling factors, or correspondingly, lower entropy for a dipolar gas in a lattice.

We thank P. Julienne for the calculation based on the two-channel model of the Feshbach resonance and S. Kotochigova for the calculation of the complex polarizability of KRb ground-state molecules.  We thank G. Qu\'em\'ener and J. L. Bohn for stimulating discussions and also their estimates of the on-site bimolecular loss rate. We gratefully acknowledge financial support for this work from NIST, NSF, AFOSR-MURI, DOE, and DARPA. S.A.M. acknowledges funding from the NDSEG Graduate Fellowship.

\normalsize{$^\ast$These authors contributed equally to this work.}\\
\normalsize{$^\dagger$To whom correspondence should be addressed; }\\
\normalsize{E-mail: Jin@jilau1.colorado.edu; Ye@jila.colorado.edu}


\end{document}


\title{\emph{Supplementary Material}

Long-lived dipolar molecules and Feshbach molecules in a 3D optical lattice}

\author{Amodsen Chotia,$^{\ast}$ Brian Neyenhuis,$^{\ast}$ Steven A. Moses, Bo Yan, Jacob P. Covey, Michael Foss-Feig, Ana Maria Rey, Deborah S. Jin,$^\dagger$ and Jun Ye$^\dagger$}
\affiliation{
\normalsize{JILA, National Institute of Standards and Technology and University of Colorado, Department of Physics, University of Colorado, Boulder, CO 80309-0440, USA}
}

\setcounter{figure}{0}
\renewcommand{\thefigure}{S\arabic{figure}}

\setcounter{equation}{0}
\renewcommand{\theequation}{S\arabic{equation}}

\maketitle
\date{today}

\section{Loss rates from ground-state molecule collisions in a corrugated 1D tube}
It was observed experimentally that the lifetime of molecules in an array of 1D tubes is sensitive to the strength of an applied electric field. This electric field dependence is expected for collisions between identical dipolar fermions \cite{ni}, and hence implicates molecule-molecule collisions as the loss mechanism.  It was also observed that, at zero electric field, the addition of a weak corrugating lattice along the tubes greatly enhances the molecule lifetime.  In order to understand this effect, here we study lossy molecule-molecule collisions in a tube, at zero electric field and in the presence of a lattice along the $z$ direction.

\begin{figure}[!h]
\centering
\subfiguretopcaptrue

\subfigure[][]{
\includegraphics[height=3.3cm]{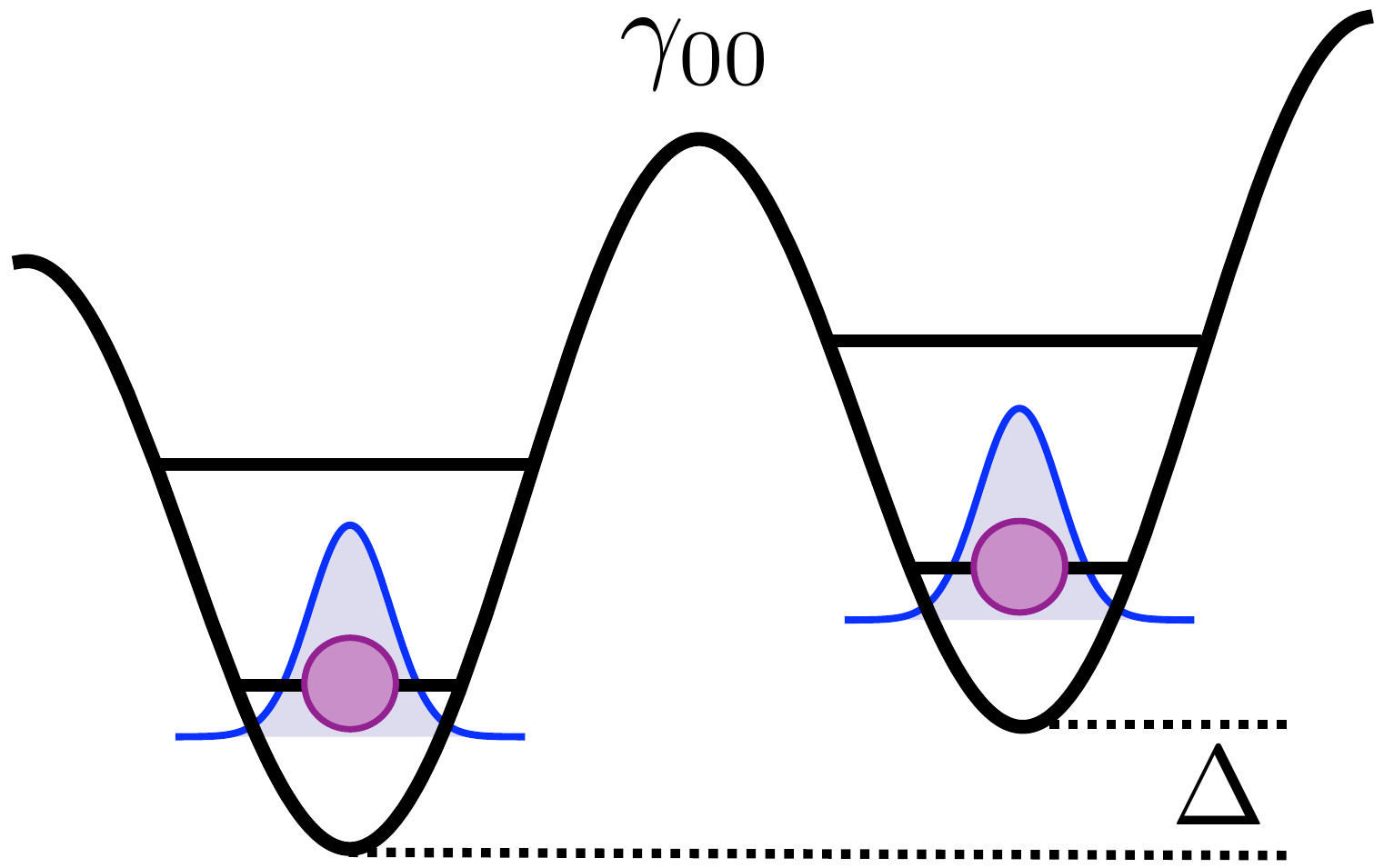}
\label{Fig1aS}}
\subfigure[][]{
\includegraphics[height = 3.3cm]{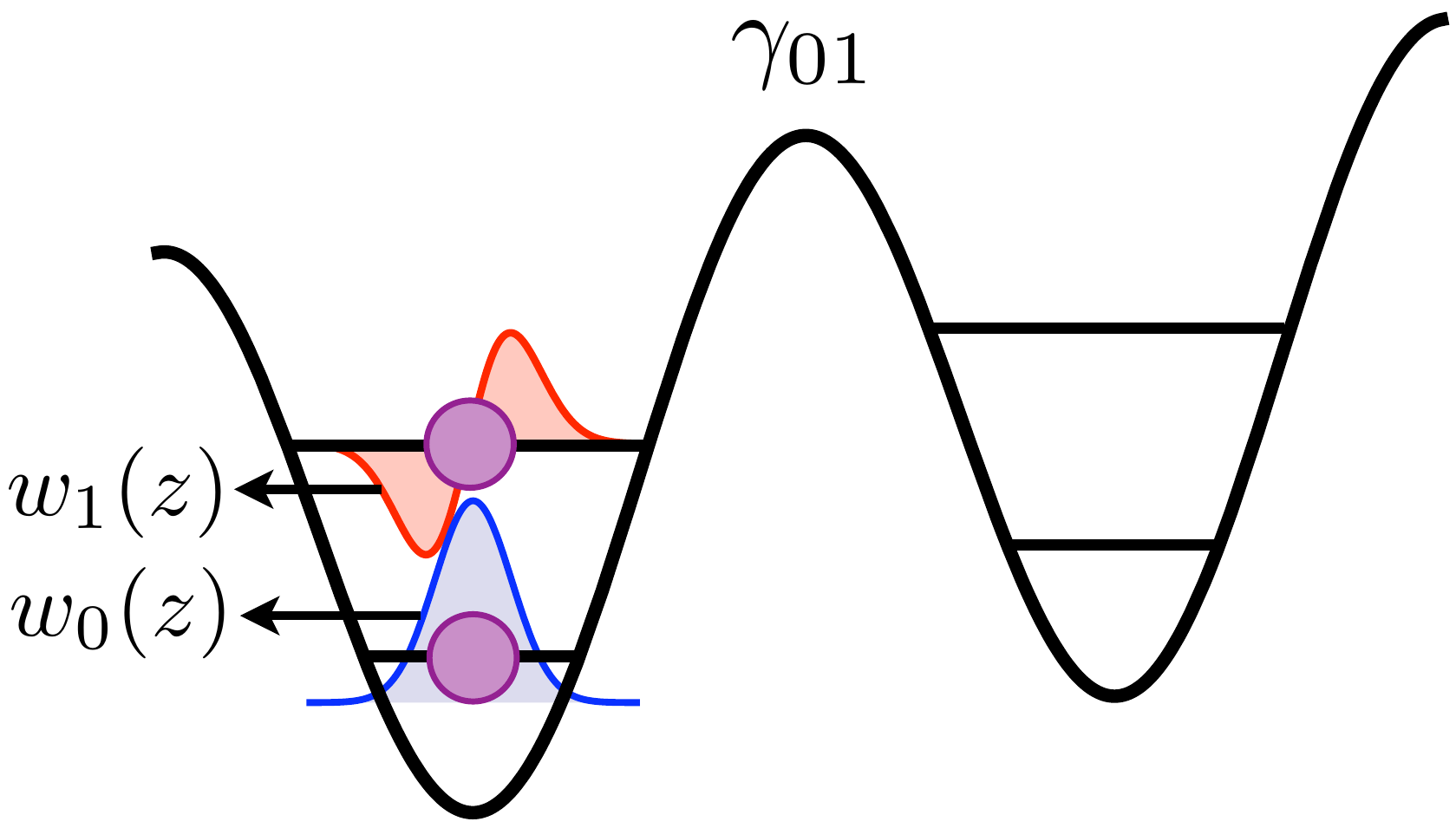}
\label{Fig1bS}}
\caption{Loss processes within the lowest two bands of the $z$ lattice.  (a) Lowest band loss rate for two molecules on neighboring lattice sites ($\gamma_{00}$). (b) Onsite loss rate for one molecule in the lowest band and one in the first excited-band on the same site ($\gamma_{01}$).  The offset $\Delta$ is in general present, due to the gaussian profile of the lattice beams.}
\label{Fig1S}
\end{figure}

At low temperatures $(\mu\mathrm{K})$ and in zero electric field, a collision between two ground-state fermionic molecules in the same nuclear spin state occurs primarily in the $p$-wave channel.  We model the lossy part of the molecule-molecule interaction with a $p$-wave pseudopotential, parametrized by a strictly imaginary scattering volume \cite{blume,julienne}.  The magnitude of the scattering volume, $V_p\approx350~\mathrm{nm}^3$, is obtained by comparing ab initio calculations of the loss rate for two molecules in an isotropic 3D harmonic trap \cite{goulven} to results using the pseudopotential.  Assuming that most molecules are in the transverse ground-state of a tube, two representative loss processes for molecules in the same tube (on two adjacent lattice sites) are shown in Fig. \ref{Fig1S}.  The loss rate for two nearest-neighbor molecules within the lowest band ($\gamma_{00}$) is rapidly suppressed by Pauli blocking for increasing $z$-lattice depth.  The loss rate for one molecule in the lowest band and one molecule in the first excited band on the same site ($\gamma_{01}$) grows with increasing $z$-lattice depth, due to the increasing localization of the lowest band and first excited band Wannier orbitals ($w_0(z)$ and $w_1(z)$, respectively).

\begin{figure}[!h]
\centering
\includegraphics[width=12.0 cm]{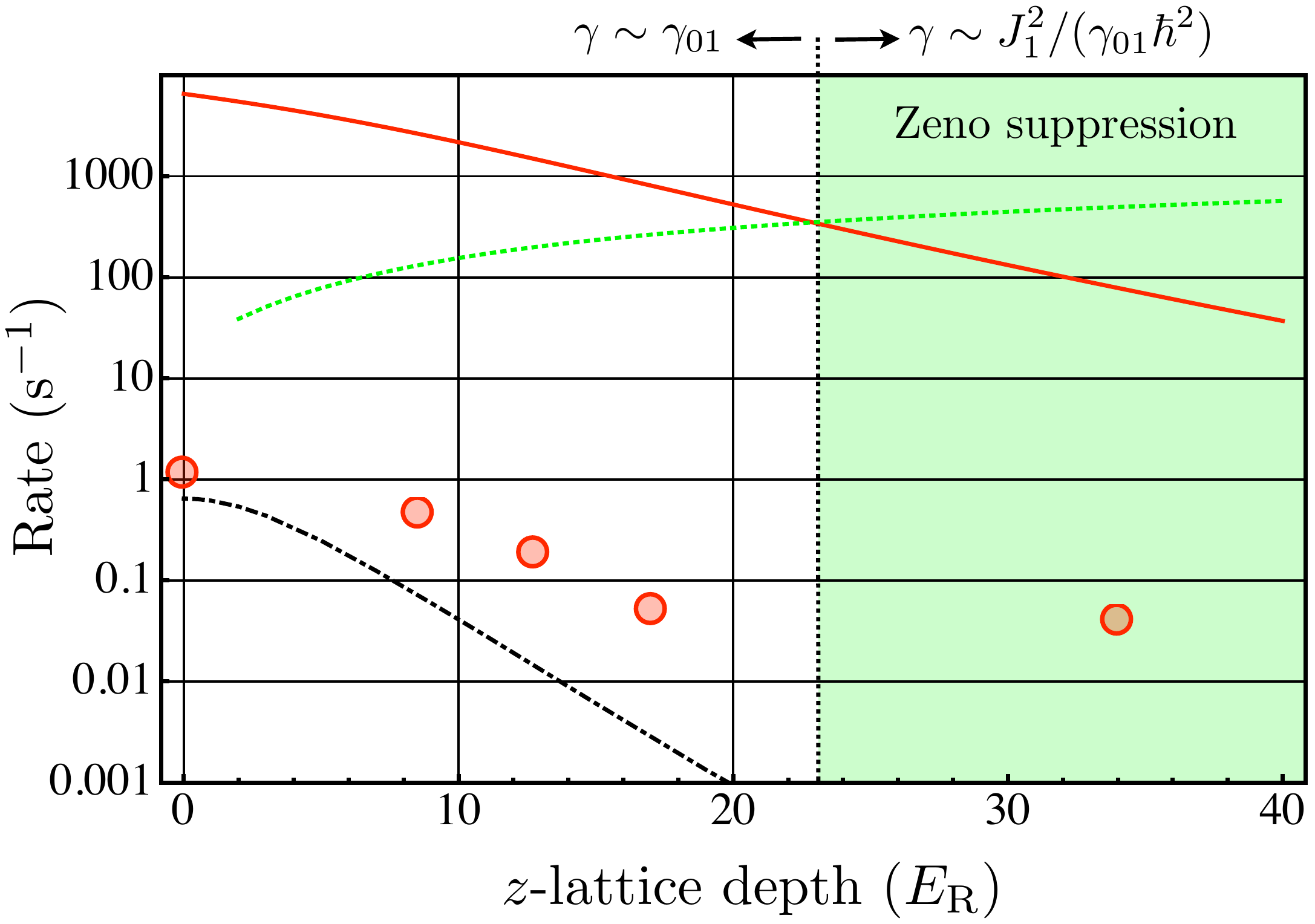}
\caption{Relevant rates for molecules confined to the lowest two bands in the $z$-direction, as a function of the $z$-lattice depth.  The black dashed-dotted curve is $\gamma_{00}$, the red solid curve is $J_1/\hbar$, and the green dotted curve is $\gamma_{01}$.  It is clear that $\gamma_{00}$ is too small to explain the measured loss rates (red circles, the error bars are smaller than the plot markers).
The rate $\gamma_{01}$ increases with the $z$-lattice depth due to the increasing confinement of the lowest two Wannier orbitals, and eventually becomes larger than the first excited-band hopping rate ($J_1/\hbar$) at a $z$-lattice depth of around 23 $E_{\mathrm{R}}$.  For even larger $z$-lattice depths (green shaded region) one expects Zeno suppression to affect the loss rate of two molecules in different bands.}
\label{Fig2S}
\end{figure}

We now address the question of whether the measured loss rates can be understood in terms of the fundamental two-body loss processes shown in Fig. \ref{Fig1S}, and their associated rates $\gamma_{00}$ and $\gamma_{01}$.  Two molecules in the lowest band should be lost at a rate proportional to $\gamma_{00}$, where the constant of proportionality is determined in a semiclassical picture by the fraction of time those two molecules spend on neighboring sites.  Since this fraction is clearly less than one---being driven down both by low density and by single particle localization due to the energy offsets between sites ($\Delta$)---$\gamma_{00}$ is an upper bound to the actual loss rate of two molecules in the lowest band.  As shown in Fig. \ref{Fig2S}, $\gamma_{00}$ is too small to explain the measured loss rates, so collisions between molecules within the lowest band are clearly not the loss mechanism.  When one molecule is in the lowest band and one molecule is in the first excited band, the loss rate is proportional to $\gamma_{01}$, so long as $\hbar\gamma_{01}\ll J_1,\Delta$ is satisfied.  It is still possible, however, that the loss rate can decrease with increasing lattice depths, since even as $\gamma_{01}$ is becoming larger the molecules are becoming more localized.  When  $\hbar\gamma_{01}\gg J_1,\Delta$, in the so-called Zeno suppression regime, the loss rate can actually be inversely proportional to $\gamma_{01}$.  However, for the experimental data taken at low $z$-lattice depths (the first four data points in Fig. \ref{Fig2S}), the Zeno suppression scenario can be ruled out, and interband collisions should result in a loss rate that is proportional to $\gamma_{01}$.

This result is, however, inconsistent with the experimental observation that losses in corrugated tubes do not exhibit the electric field dependence of losses in uncorrugated tubes: When dipole-dipole interactions are taken into account, both $\gamma_{00}$ and $\gamma_{01}$ should increase with electric field strength.  The most likely explanation for this lack of electric field dependence is that, in the presence of the $z$-lattice, collisional physics does not determine the molecule lifetime, despite being ultimately responsible for the loss.  Instead, a heating mechanism that promotes molecules from localized to mobile single-particle levels is what sets the time scale for loss.  If the promotion rate into mobile states is slower than the collisional loss rate for mobile molecules, but faster than the collisional loss rate for localized molecules---which is suppressed by the diluteness of the molecular gas---then it will determine the molecule lifetime, thereby rendering the lifetime independent of electric field.

To model this process phenomenologically, we assume that a heating mechanism imparts energy to the molecules at a fixed rate, and hence the promotion rate between successive bands is inversely proportional to the band gap between them.  Microscopically, this picture is motivated by the following reasoning.  Fluctuations in the transverse lattice beams---which provide a potential that is only approximately decoupled from the $z$-direction---should be the dominant source of heating at low $z$-lattice depths.  While a detailed knowledge of these fluctuations is absent, one can in complete generality assume that to first order in $z$, the motion of a molecule along $z$ couples to a time-dependent perturbation of the form $\mathcal{I}(z,t)=\mathcal{I}_{\perp}f(t)z$, where the intensity $\mathcal{I}_{\perp}$ depends primarily on the transverse lattice strength.  The resulting transition rate from a localized state $|\alpha\rangle$ (Wannier orbital of the $\alpha^{\mathrm{th}}$ band along $z$) to a localized state $|\alpha+1\rangle$ (Wannier orbital of the next higher band along $z$) on the same lattice site is given by \cite{thomas}
\begin{equation}
\Gamma_{\alpha,\alpha+1}=\frac{\pi S(\omega_{\alpha})\mathcal{I}_{\perp}^2}{\hbar^2}|\langle \alpha+1|z|\alpha\rangle|^2.
\end{equation}
In the above, $S(\omega_{\alpha})$ is the power spectrum of the noise $f(t)$, evaluated at the transition frequency $\omega_{\alpha}$ between $|\alpha\rangle$ and $|\alpha+1\rangle$.  For a power spectrum that is relatively flat over frequency ranges $\Delta\omega\sim E_R/\hbar$, the matrix element $|\langle \alpha+1|z|\alpha\rangle|^2\sim\omega_{\alpha}^{-1}$ gives the dominant scaling of the transition rate on the $z$-lattice depth, and thus we expect interband transitions of this variety to be suppressed by the band gaps.  On a single site, if we take the energies of localized states to be the band centers $\epsilon_{\alpha}(v)$ (where $v$ is the $z$-lattice depth measured in $E_R$), then we can define interband transition rates $\Gamma_{\alpha,\alpha+1}(v)=\mathcal{E}[\epsilon_{\alpha+1}(v)-\epsilon_{\alpha}(v)]^{-1}$, where $\mathcal{E}$ is a constant heating rate.

The loss rate for a given molecule is obtained by deciding on a localization criterion; for a molecule in the $\alpha^{\mathrm{th}}$ band sitting $s$ sites from the trap center (in the $z$-direction), we consider that it is mobile whenever the bandwidth $4J_{\alpha}$ provides enough kinetic energy for it to reach the trap center \cite{hooley}.  This criterion sets a critical lattice depth $v_c(\alpha,s)$ in the $\alpha^{\mathrm{th}}$ band, below which a molecule at site $s$ is mobile and above which it is localized.  Our primary assumption is that the decay rate for a mobile molecule is much larger than the interband transition rate, in which case it can be chosen to be infinite.  We then obtain the loss rate for a molecule on a given site and in a given band by solving coupled rate equations that describe the heating and loss.  Because the experiments measure a molecule lifetime that is averaged over many molecules dispersed throughout the trap, we assume an initial distribution of molecules in the $z$-direction that is gaussian with a $1/e^2$ radius of $\xi=14$ lattice sites, which is based on experimental measurements of the molecule cloud, and average the results of our calculations over this distribution.  The qualitative features of the results are not particularly sensitive to $\xi$.  We find that heating rates between $\mathcal{E}=1~E_R/s~(66~\mathrm{nK}/s)$ and $\mathcal{E}=2~E_R/s~(132~\mathrm{nK}/s)$ are consistent with the experimental data at low $z$-lattice depths, and molecule lifetimes as functions of the $z$-lattice depth for these two heating rates are shown in Fig. 2 of the manuscript.  It should be noted that once the $z$ lattice is sufficiently deep, and molecules remain localized for times comparable to the inelastic light scattering time, this model is no longer valid.
